\documentclass[onecolumn,numberedappendix]{revtex4-2}

\usepackage{amsmath, amssymb, amsfonts}
\usepackage{graphicx}

\usepackage[colorlinks=true, 
            linkcolor=blue, 
            urlcolor=blue, 
            citecolor=magenta]{hyperref}

\usepackage{orcidlink}

\usepackage{xcolor} % Required package
\usepackage{float} % For [H] float option

\begin{document}

%%%%%%%%%%%%%%%%%%%%%%%%%%%%%%%%%%%%%%%%%%%%%%%%%%%%%%%%%%%%%%%%%%%%%%
\title{Tests of general relativity in pseudo-Newtonian approach}%

\author{Naman Goyal}\thanks{namangoyal@iisc.ac.in}
\affiliation{Department of Physics, Indian Institute of Science, Bangalore 560012}

\author{Banibrata Mukhopadhyay}\thanks{bm@iisc.ac.in}
\affiliation{Department of Physics, Indian Institute of Science, Bangalore 560012}

\author{Ashish Kumar Meena}\thanks{akm@iisc.ac.in}
\affiliation{Department of Physics, Indian Institute of Science, Bangalore 560012}

%%%%%%%%%%%%%%%%%%%%%%%%%%%%%%%%%%%%%%%%%%%%%%%%%%%%%%%%%%%%%%%%%%%%%%
\begin{abstract}
We investigate the extent to which pseudo-Newtonian gravitational potentials can reproduce classic tests of general relativity without resorting to full general relativistic formalisms. This is useful for the researchers seeking intuitive insight into relativistic gravity. Focusing on the perihelion precession of Mercury, gravitational redshift, and gravitational light bending, we derive analytical expressions for orbital precession and demonstrate that, with suitable physically acceptable parameters, pseudo-Newtonian approaches can accurately reproduce the observed perihelion advance and gravitational redshift. However, we confirm that no single potential consistently captures all relativistic effects. In particular, while certain parameters yield agreement with general relativity for planetary motion and redshift, they fail to reproduce gravitational lensing over a broad range of impact parameters. Our results highlight both the usefulness and limitations of pseudo-Newtonian methods in modeling gravitational phenomena. Although the pseudo-Newtonian approach cannot serve as universal substitutes for general relativity, especially in strong-field regimes, it provides valuable semi-analytical insight and pedagogical simplicity. Our results indicate the usefulness of the pseudo-Newtonian approach to uncover more complicated phenomena involved with strong field gravity in possibly modification to general relativity. 
\end{abstract}

%%%%%%%%%%%%%%%%%%%%%%%%%%%%%%%%%%%%%%%%%%%%%%%%%%%%%%%%%%%%%%%%%%%%%%
\maketitle

%%%%%%%%%%%%%%%%%%%%%%%%%%%%%%%%%%%%%%%%%%%%%%%%%%%%%%%%%%%%%%%%%%%%%%
\section{Introduction}
Einstein's General Theory of Relativity~(GR) \cite{einstein15} is one of the most celebrated discoveries of the 20th century. The three classic tests of GR include the perihelion precession of Mercury, the deflection of light (also known as gravitational lensing) by the Sun, and the gravitational redshift. The first test that GR passed was the prediction of the correct value for Mercury's perihelion precession \cite{einstein15}. In 1919, the deflection of light by the Sun was also put to the test, and the observations agreed with the GR predictions \cite{edd20}. The last classic test of GR, i.e., gravitational redshift, was put to the test in the famous Pound-Rebka experiment \cite{pound60}, again agreeing with the GR predictions. Additional experimental/observational tests were proposed (including the Shapiro delay \cite{shapiro64} and the presence of gravitational waves \cite{ligo16}), and the results agree well with the GR predictions, making it the most successful theory of gravity at present.

Due to the non-linear nature of GR, it often becomes challenging to exactly solve the underlying equations, particularly when the spacetime is deviated from the spherically symmetery, i.e., due to rotating central objects. Hence, theoretical studies often make use of a pseudo-Newtonian approach, where the dynamics is described by Newtonian equations, but the underlying gravitational potential is modified in such a way that it captures certain basic GR effects. This is especially very popular and successful for understanding the accretion disk around a compact object, where one needs to deal with general relativistic magnetohydrodynamical (GRMHD) flows. For example, Paczy\'nsky \& Wiita~\cite{pw80} introduced a pseudo-Netonian potential that can correctly reproduce marginally bound and marginally stable orbits for the Schwarzschild metric. Later, other pseudo-Newtonian potentials by, e.g., Chakrabarti \& Khanna \cite{ck92}, Artemova et al. \cite{abn96}, Mukhopadhyay \cite{bm02}, Mukhopadhyay \& Misra \cite{mm03}, were proposed to study accretion disk around a rotating black hole. In fact, Mukhopadhyay \cite{bm02} proposed a first principle based method to derive a pseudo-Newtonian potential for the Keplerian motion in any spacetime metric. These days however, thanks to advanced techniques and increased computational power, researchers do perform GRMHD simulations~\citep[e.g.,][]{gammie03, mckinney07, pathak25}. Nonetheless, the errors generated in the pseudo-Newtonian approach are often smaller than the observational/instrumental uncertainties, depending on the problem under consideration. Therefore, on many occasions, for all practical purposes, the pseudo-Newtonian approach suffices. More so, in order to understand the underlying basics, often the (semi)analytical approach is very useful, where the usefulness of pseudo-Newtonian approaches comes in. That said, it is important to emphasize that there is no real alternative to the pure GR at present.

In our current work, we subject the three classic tests of GR, i.e., perihelion precession of Mercury, gravitational redshift, and gravitational lensing, to different pseudo-Newtonian potentials. This has two purposes: (1) To verify how useful the pseudo-Newtonian approaches are globally and how accurately they are able to capture well-known GR properties, which would help assess their applicability to the complicated problems whose exact results are not known yet. (2) To let the difficult but fundamental GR concepts reach the academicians of all branches, particularly of Physics, who may not even be experts in GR. We show that depending on the pseudo-Newtonian potential and the underlying parameters, we are able to obtain the Mercury's perihelion precession and gravitational redshift exactly the same as the GR values. However, the pseudo-Newtonian approach does not seem to be adequate for gravitational lensing.

The plan of the paper is the following. In the next section, \S\ref{sec:pots}, we recapitulate some important pseudo-Newtonian potentials. Subsequently, in \S\ref{sec:peri} we formulate the basic equations for the precession of a planet (or motion of any body) in an orbit around a central gravitating body and obtain the precession frequency of Mercury for the Paczy\'nsky-Wiita (PW) \cite{pw80} and Artemova-Bjornson-Novikov (ABN) \cite{abn96} potentials. Note that as the radius of Sun is several orders of magnitude larger than its gravitational radius, the GR effect due to solar rotation is practically zero and, hence, the potential for the spherically symmetric spacetime suffices. For the same reason, a pseudo-Newtonian potential for black hole works for Sun. We further explore gravitational redshift and lensing in the pseudo-Newtonian approach in \S\ref{sec:redshift} and \S\ref{sec:deflection}, respectively, before we conclude in \S\ref{sec:conclusion}.

%%%%%%%%%%%%%%%%%%%%%%%%%%%%%%%%%%%%%%%%%%%%%%%%%%%%%%%%%%%%%%%%%%%%%%
\section{Recapitulation of Gravitational Potentials for the problem}
\label{sec:pots}

Before we describe the potential, let us mention the units of the system.
We consider the units of length $r_g=GM/c^2$, velocity $c$ and specific angular momentum $GM/c$, where $G$ and $M$ are respectively Newton's gravitation constant and mass of the central object, and $c$ is the speed of light. Below we recapitulate the useful pseudo-Newtonian potentials.

%%%%%%%%%%%%%%%%%%%%%%%%%%%%%%%%%%%%%%%%%%%%%%%%%%%%%%%%%%%%%%%%%%%%%%
\subsection{Mukhopadhyay potential}

The force corresponding to the pseudo-Newtonian potential proposed by Mukhopadhyay \cite{bm02} is given by,
\begin{eqnarray}
F=-\frac{dV(x)}{dx} = -\frac{ (x^2 - 2a \sqrt{x} + a^2)^2}{x^3 \left( \sqrt{x}(x-2) + a \right)^2},
\label{bm}
\end{eqnarray}
where $x$ is the radial coordinate from the black hole (or the source of gravity) and $a$ is the specific angular momentum of the black hole. For $a=0$, it reduces to the PW potential \cite{pw80}, given by
\begin{eqnarray}
V=-\frac{1}{x-2}.
\label{pw}
\end{eqnarray}

For the present purpose, neglecting $a$ is good enough, as we discuss physics away from the source of gravity, e.g. at the 
surface of Sun where $x\simeq4.7\times10^5$, whereas $|a|\le 1$.

%%%%%%%%%%%%%%%%%%%%%%%%%%%%%%%%%%%%%%%%%%%%%%%%%%%%%%%%%%%%%%%%%%%%%%
\subsection{ABN potential}

Another pseudo-Newtonian potential (and corresponding force), namely the ABN potential~\cite{abn96}, is given by,
\begin{equation}
F=-\frac{dV}{dx} =-\frac{1}{x^{2-\beta}(x-x_{H})^\beta},
\label{artemova}
\end{equation}
where $x_{H}$ is the event horizon radius. We further amend the potential for the present purpose, considering $\beta$ to be a parameter, unlike the original proposal, when it was constrained to be a function of black hole spin,~$a$.    
As mentioned above, for the present purpose, the spin effect of the compact object can be neglected and thence $x_{H}=2$. Also, 
$\beta=2$ reduces $V$ to the PW potential and $\beta=0$ to the Newtonian potential.

%%%%%%%%%%%%%%%%%%%%%%%%%%%%%%%%%%%%%%%%%%%%%%%%%%%%%%%%%%%%%%%%%%%%%%
\section{Perihelion precession of Mercury}
\label{sec:peri}
 
The total energy per unit mass, $E$, and the angular momentum per unit mass, $l$, of Mercury are given by,
\begin{align}
E &= \frac{\left[ \dot{x}^2 + (x \dot{\theta})^2 \right]}{2} + V(x), \\
l &= x^2 \dot{\theta},
\end{align}
where $\theta$ is the angular coordinate measured from the center of the Sun (in general, the source of gravity). Therefore,
\begin{equation}
\frac{\dot{x}^2}{2} = E - V(x) - \frac{l^2}{2x^2}.
\label{xdot1}
\end{equation}

Now expressing  $\dot{x}$ in terms of $\theta$ as
$\dot{x} = \dot{\theta}dx/d\theta  = l(dx/d\theta)/x^2$
and substituting into Eq.~(\ref{xdot1}) we obtain 
\begin{equation}
\frac{l^2}{2x^4} \left( \frac{dx}{d\theta} \right)^2 = E - V(x) - \frac{l^2}{2x^2}.
\label{xdot2}
\end{equation}

Further, substituting $\alpha = 1/x$, hence
$dx/d\theta = -1/\alpha^2 (d\alpha/d\theta)$ in Eq.~(\ref{xdot2}),
we obtain
\begin{equation}
\frac{l^2}{2} \left( \frac{d\alpha}{d\theta} \right)^2 = E - V(\alpha) - \frac{l^2 \alpha^2}{2}.
\label{xdot3}
\end{equation}

Differentiating Eq.~(\ref{xdot3}) with respect to $\theta$ and dividing by $d\alpha/d\theta$ (assuming $\neq 0$), we obtain the general orbit equation as
\begin{eqnarray}
l^2 \frac{d^2\alpha}{d\theta^2} = -\frac{dV}{d\alpha} - l^2 \alpha.
\label{orbit1}
\end{eqnarray}

\subsection{Solution with PW potential}
\label{pwsol}

For the PW potential $V(\alpha) = -\alpha/(1-2\alpha)$, and we obtain the equation for an orbit as
\begin{equation}
\frac{d^2\alpha}{d\theta^2} = -\alpha + \frac{1}{l^2 (1-2\alpha)^2}.
\label{orbit2}
\end{equation}
Further setting $k = 1/l^2$ and $\lambda = \alpha/k$, Eq.~(\ref{orbit2}) reduces to
\begin{equation}
\frac{d^2\lambda}{d\theta^2} = -\lambda + \frac{1}{(1 - 2\lambda k)^2}.
\label{orbit3}
\end{equation}
Assuming $\lambda k \ll 1$ (which is justified for the present purpose, verified later), the orbit equation reduces to
\begin{equation}
\frac{d^2\lambda}{d\theta^2} + (1 - 4k)\lambda = 1,
\label{orbit4}
\end{equation}
whose solution is given by
\begin{equation}
\alpha = k \left[ A \cos \left( \sqrt{1 - 4k} \, \theta + \phi \right) + \frac{1}{1 - 4k} \right].
\label{orbit4}
\end{equation}
In the Newtonian limit ($x \gg 2$), the form of the solution for an orbit is 
\begin{eqnarray}
\alpha = \frac{1}{R} (1 + \varepsilon \cos \theta), 
\label{Newt}
\end{eqnarray}
where $R$ is the dimensionless semi-major axis and $\varepsilon$ is the eccentricity of the orbit. Comparing Eq.~(\ref{Newt}) and Eq.~(\ref{orbit4}), we
interpret
\begin{align*}
k &\approx \frac{1}{R} \quad (\text{for } R \gg 4), \\
A &\approx \varepsilon, \\
\phi &= 0 \quad (\text{initial phase}).
\end{align*}
Thus, Eq. (\ref{orbit4}) becomes
\begin{eqnarray}
\alpha \approx \frac{1}{R} \left[ \varepsilon \cos \left( \sqrt{1 - 4/R} \, \theta \right) + \frac{R}{R - 4} \right].
\label{orbsol1}
\end{eqnarray}

\subsubsection{Orbital Precession}
\label{pwsol1}

The argument of cosine gives the precession per orbit. The angular advance per orbit is
\[
\Delta \theta = 2\pi \left( \frac{1}{\sqrt{1 - 4k}} - 1 \right) \approx 2\pi \left( \frac{1}{1 - 2k} - 1 \right) \approx 4\pi k \quad \text{radians/orbit},
\]
when \( \sqrt{1 - 4k} \approx 1 - 2k \) and \( (1 - 2k)^{-1} \approx 1 + 2k \). Substituting \( k \approx {1}/{R} \), we obtain
\[
\Delta \theta \approx \frac{4\pi}{R} \quad \text{radians/orbit}.
\]

\subsubsection{Precession per Century}
\label{pw-prec}
Mercury's parameters:
\begin{itemize}
\item Semi-major axis: 
%\remove{\( R_d = 5.55 \times 10^{10} \, \text{m} \)} 
\( R_d = 5.79\times 10^{10} \, \text{m}\)
\item 
%\remove{\( r_g = GM_\odot/c^2 = 1474.14 \, \text{m} \)} 
\( r_g = GM_\odot/c^2 = 1476.62 \, \text{m} \)
\item 
%\remove{\( R = R_d/r_g \approx 3.76 \times 10^7 \)} 
\( R = R_d/r_g = 3.92 \times 10^7 \)
\item Orbital period: \( T = 0.24 \, \text{Earth years} \)
\end{itemize}
Precession per orbit is then
\[
%\Delta \theta \approx 4\pi k, \quad k = \frac{r_g}{a} = \frac{1}{R} \approx 2.66 \times 10^{-8} \, \text{radians/orbit}.
\Delta \theta \approx \frac{4\pi}{R} %\remove{\approx 3.34 \times 10^{-7}} 
=\:3.20 \times 10^{-7} \, \text{radians/orbit}.
\]
Converting to arcseconds and computing per century, we obtain
\[
\Delta \theta_{\text{century}} = \Delta \theta \times \frac{100}{T} \times \frac{180 \times 3600}{\pi} %\remove{\approx 28.6} 
= 27.54 \, \text{arcseconds/century}.
\]

Clearly, while the result is not bad, it is about a factor of one and half smaller than that of GR.

\subsubsection{Verification of Approximation}
The maximum \( \alpha \) (at perihelion) is
\[
\alpha_{\text{max}} = \frac{r_g}{R_d(1 - e)} \approx \frac{1476.62}{4.60 \times 10^{10}} \approx 3.21 \times 10^{-8}.
\]
Therefore,
\[
\lambda k = \alpha \leq \alpha_{\text{max}}  \approx 3.36 \times 10^{-8}   \ll 1.
\]

\subsection{Solution with ABN potential }
\label{abnsol}

Now we introduce the amended ABN potential given by Eq. (\ref{artemova}) to  Eq. (\ref{orbit1}) and the equation for an orbit 
%under this general potential is found by first calculating the term $\frac{dV}{d\alpha}$:
%
%\[
%\frac{dV}{d\alpha} = \frac{dV}{dx} \frac{dx}{d\alpha} = \left( -\frac{1}{x^{2-\beta}(x-2)^\beta} \right) \left( \frac{1}{\alpha^2} \right) = -1 \frac{\alpha^{2-\beta} \alpha^\beta}{(1-2\alpha)^\beta} \frac{1}{\alpha^2} = \frac{1}{(1-2\alpha)^\beta}.
%\]
reduces to
\begin{equation}
l^2 \frac{d^2\alpha}{d\theta^2} = \frac{1}{(1 - 2\alpha)^\beta} - l^2 \alpha.
\label{orbit6}
\end{equation}
Like \S \ref{pwsol}, setting $k = 1/l^2$ and $\lambda = \alpha/k$, Eq. (\ref{orbit6}) reduces to
\begin{equation}
\frac{d^2\lambda}{d\theta^2} = -\lambda + \frac{1}{(1 - 2\lambda k)^\beta},
\label{orbit7}
\end{equation}
which for $\lambda k \ll 1$ leads to 
\begin{equation}
\frac{d^2\lambda}{d\theta^2} + (1 - 2 \beta k)\lambda = 1,
\label{orbit8}
\end{equation}
with the solution given by
\begin{equation}
\alpha = k \left[ A \cos \left(\theta \sqrt{1 - 2\beta k} + \phi \right) + \frac{1}{1 - 2 \beta k} \right].
\label{orbit9}
\end{equation}
In the Newtonian limit ($\beta=0$), the solution for an orbit 
reduces to the form of Eq. (\ref{Newt}). Like the PW potential,
%\begin{eqnarray}
%\alpha = \frac{1}{R} (1 + \varepsilon \cos \theta), 
%\label{}
%\end{eqnarray}
%where $R$ is the dimensionless semi-major axis and $\varepsilon$ is the eccentricity of the orbit. 
comparing Eq. (\ref{orbit9}) with Eq. (\ref{Newt}), we
interpret
\begin{align*}
k &\approx \frac{1}{R} \quad (\text{for } R \gg 4), \\
A &\approx \varepsilon, \\
\phi &= 0 \quad (\text{initial phase}).
\end{align*}
Thus, Eq. (\ref{orbit9}) becomes
\begin{eqnarray}
\alpha \approx \frac{1}{R} \left[ \varepsilon \cos \left(\theta\sqrt{1 - 2\beta /R}  \right) + \frac{R}{R - 2 \beta} \right].
\label{orbsol2}
\end{eqnarray}

\subsubsection{Orbital Precession}
The argument of the cosine in Eq. (\ref{orbsol2}) gives the precession per orbit. The angular advance per orbit is then
\[
\Delta \theta = 2\pi \left( \frac{1}{\sqrt{1 - 2 \beta k}} - 1 \right)  \approx 2\pi \beta k \quad \text{radians/orbit},
\]
when \( \sqrt{1 - 2\beta k} \approx 1 - \beta k \) and \( (1 - \beta k)^{-1} \approx 1 + \beta k \). Substituting \( k \approx 1/R \), we obtain
\[
\Delta \theta \approx \frac{2\pi \beta}{R} \quad \text{radians/orbit}.
\]
%\subsubsection{Precession per Century}
Based on the parameters for Mercury mentioned in \S \ref{pw-prec}, we obtain the precession per century as
\[
\Delta \theta_{\text{century}} = \Delta \theta \times \frac{100}{T} \times \frac{180 \times 3600}{\pi} \approx 13.8 \, \beta \, \text{arcseconds/century}.
\]
Clearly, $\beta = 3.125$ leads to the results obtained in exact GR. Below, we verify how the various choices of $\beta$ reproduce the fundamental properties of GR.

%\subsection{Summary of Results}
%
%\begin{table}[H]
%\centering
%\caption{Summary of Mercury's orbital precession calculations}
%\label{tab:precession_summary}
%\vspace{0.5em}
%\begin{tabular}{|c|c|c|c|}
%\hline
%\textbf{Case} & \textbf{Potential} & \textbf{Precession Formula} %& \textbf{Value (arcsec/century)} \\
%\hline
%1 & \( V(x) = -\dfrac{c^2}{x-2} \) & \( 4\pi\kappa \times \dfrac{100}{T} \times \dfrac{180 \times 3600}{\pi} \) & 28.575 \\
%\hline
%2 & \( \dfrac{dV}{dx} = \dfrac{c^2 (x^2 - 2a \sqrt{x} + a^2)^2}{x^3 \left( \sqrt{x}(x-2) + a \right)^2} \) & Same as Case 1 & 28.575 \\
%\hline
%3 & \( \dfrac{dV}{dx} = \dfrac{c^2}{x^{2-\beta}(x-2)^\beta} \) & %\( 14.29\beta \) & \( 14.29\beta \) \\
%\hline
%\end{tabular}
%\end{table}

%Where:
%\begin{itemize}
%\item \(\kappa = r_g/a = 2.6576 \times 10^{-8}\)
%\item Orbital period \(T = 0.24085\) yr
%\item \(\beta\) is the potential parameter
%\end{itemize}

\subsection{Radii of fundamental orbits}

There are fundamental characteristic orbits
appearing in GR which are absent in Newtonian dynamics. They are, e.g., last (marginally) stable and bound orbit radii (respectively denoted by $x_{\rm ms}$ and $x_{\rm mb}$), photon orbit radius.  
We know that the PW and Mukhopadhyay potentials reproduce $x_{\rm ms}$ as of GR exactly. While the Mukhopadhyay potential reproduces $x_{\rm mb}$ with a small error, the PW (which is the Mukhopadhyay potential with $a=0$) does also it exactly. Also both the potentials reproduce energy at $x_{\rm ms}$ within $10\%$ error. Below we tabulate these fundamental properties
%along with gravitational redshift 
for various values of underlying parameters and those in GR in Table \ref{tab1}.

\begin{table}[H]
\centering
\caption{Radii of marginally stable ($x_{\rm ms}$) and marginally bound ($x_{\rm mb}$) orbits for different $\beta$.}
\label{tab1}
\vspace{0.5em}

\begin{tabular}{|c|c|c|c|c|} 
\hline

$\beta$ & $x_{\text{ms}}$ & $x_{\text{mb}}$ & $\Delta\theta$ (arcsec/century) & $E_{\text{ms}}$  \\
\hline
0       & 0              & 0              & 0                  & - \\
0.5       & 3              & 2.25              & 6.89             & -0.13397 \\
1.5       & 5              & 3.39              & 20.66             & -0.07583\\
2       & 6              & 4              & 27.54             & -0.06249 \\
2.25    & 6.5            & 4.31          & 30.99             & -0.05746 \\
3.125  & 8.25         & 5.39          & 43.03           & -0.04476 \\
\hline
\hline
GR & 6 & 4 & 43.03  & -0.05719 \\
for $a=0$ & & & & \\
\hline
\end{tabular}
\end{table}

Table \ref{tab1} confirms that $\beta = 3.125$ yields a perihelion precession of Mercury consistent with the observed value. However, for that $\beta$, $x_{\rm ms}$ and $x_{\rm mb}$ increase to $37.5\%$. However, as the solar radius is significantly larger than 
$x_{\rm ms}$ and $x_{\rm mb}$ for Sun, this does not pose any practical problem for the present purpose. Therefore, for all the practical purposes, an ABM potential amended by us can be used for computing precession of planetary orbits in the solar system.

%\section{Conclusions}

%The analysis demonstrates consistent orbital precession predictions for Mercury across different potential formulations:

%\begin{enumerate}
%\item \textbf{Case 1 and 2 equivalence}: Both potentials yield identical precession (28.575 arcsec/century), confirming that the specific form of the $a$-dependent terms in Case 2 does not affect the first-order orbital dynamics.

%\item \textbf{General potential (Case 3)}: The precession follows \(P = 14.29\beta\) arcsec/century, providing:
%\begin{align*}
%P &= 28.575 \text{ arcsec/century} & \text{when } \beta &= 2 \\
%P &= 43.01 \text{ arcsec/century} & \text{when } \beta &= 3.012
%\end{align*}

%The first-order approximation remains valid for Mercury's orbital parameters, with maximum \(\lambda k \approx 8.9 \times 10^{-16} \ll 1\). The general potential formulation successfully unifies the different cases through the \(\beta\) parameter.

%\newpage

%%%%%%%%%%%%%%%%%%%%%%%%%%%%%%%%%%%%%%%%%%%%%%%%%%%%%%%%%%%%%%%%%%%%%%
\section{Gravitational Redshift}
\label{sec:redshift}

Gravitational redshift ($z$) is a measure of the energy lost by a photon as it travels out of a gravitational field. It is defined as the fractional change in wavelength observed ($\lambda_{\rm obs}$) with respect to that emitted ($\lambda_{\rm emit}$) as $z = (\lambda_{\rm obs} - \lambda_{\rm emit}) / \lambda_{\rm emit}$. We compare $z$ obtained based on the analytical GR based model with that based on several pseudo-Newtonian potentials.

\begin{figure}[h!] 
    \centering
    \includegraphics[width=1\textwidth]{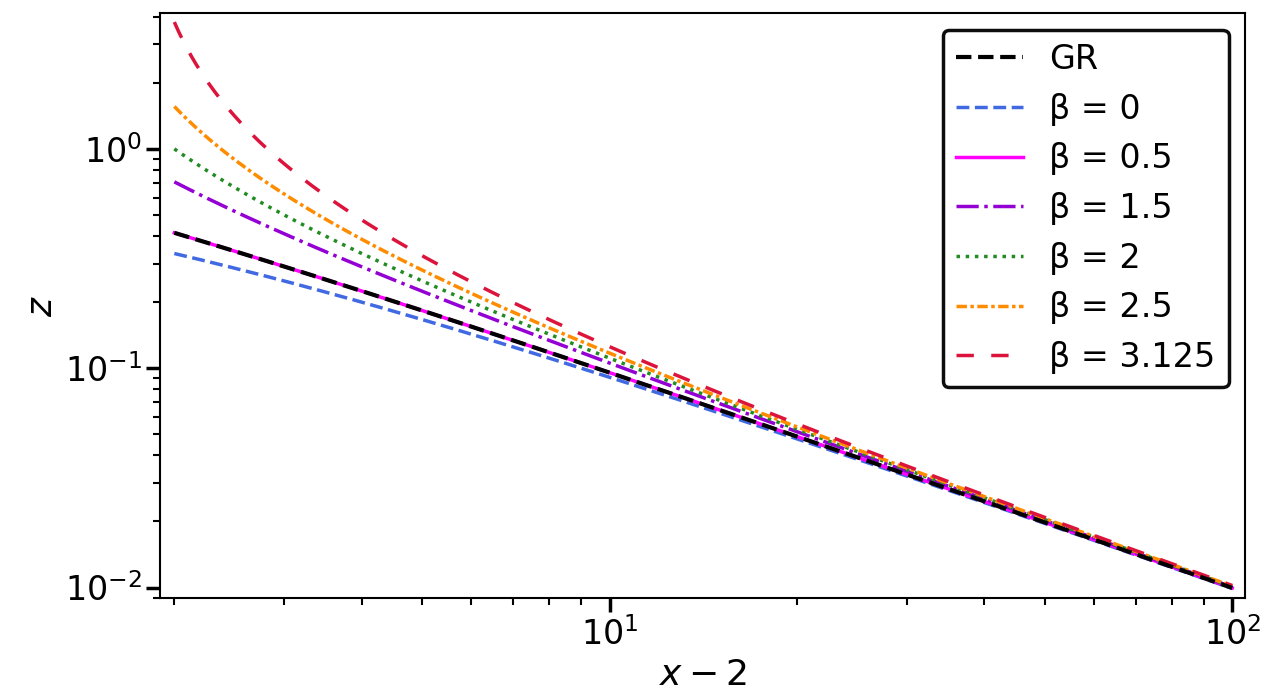}
    \caption{Variation of gravitational redshift as a function of radial coordinate for the ABN pseudo-Newtonian potential with different $\beta$. The exact solution from GR (black dashed) is also shown which matches with high accuracy for $\beta=0.5$ (solid magenta). All models converge at large distances, as expected, but diverge significantly in the strong-field regime near the black hole horizon.}
    \label{fig:redshift_log_plot}
\end{figure}

For the case of the GR of a nonrotating compact object, e.g. black hole, $z$ for a photon emitted at a radius $x$ and observed at infinity is given by 
\begin{equation}
    z_{GR}(x) = \frac{1}{\sqrt{1 - \frac{2}{x}}} - 1.
\end{equation}
For the pseudo-Newtonian potential, $z$ is however calculated by integrating the potential gradient from the source at radius $x$ to the observer at infinity, given by
\begin{equation}
    z_{\text{pseudo}}(x; \beta) = \frac{-V(x)}{1+V(x)},
    \label{eq:redshift_pseudo}
\end{equation}
where $V$ follows from Eq. (\ref{pw}) or Eq. (\ref{artemova}). Fig. \ref{fig:redshift_log_plot} shows how $z$ varies with the distance from the vicinity of a black hole for various $\beta$ of the ABN potential. Interestingly, it deviates significantly from that of GR for $\beta=3.125$ which otherwise leads to the exact precession of Mercury. Nevertheless, $\beta=0.5$ matches the GR result exactly. 

Therefore, while pseudo-Newtonian potential is a very useful tool to capture GR phenomena, at least approximately, without going through rigorous GR, there does not seem to have a unique potential for all the purposes. For the accretion disk around a non-rotating black hole $\beta=2$ suffices very nicely, for the perihelion precession of Mercury it is $\beta\approx 3$, while for gravitational redshift $\beta=0.5$ does the job. Moreover, for epicyclic oscillation frequencies, none of these $\beta$ seem to be good, one needs a different pseudo-Newtonian approach \cite{nw91,mm03}. Nevertheless, far away from the event horizon, e.g. for Sun, the results from GR, pseudo-Newtonian and Newtonian approaches match, as is clear from Fig. \ref{fig:redshift_log_plot}.

%%%%%%%%%%%%%%%%%%%%%%%%%%%%%%%%%%%%%%%%%%%%%%%%%%%%%%%%%%%%%%%%%%%%%%

%%%%%%%%%%%%%%%%%%%%%%%%%%%%%%%%%%%%%%%%%%%%%%%%%%%%%%%%%%%%%%%%%%%%%%
\section{Gravitational Lensing}
\label{sec:deflection}
Gravitational lensing refers to the bending/deflection of light rays as they pass through the gravitational field of a massive object. It is quantified by the total deflection angle, $\delta$, which measures the deviation of the photon's path from its original trajectory as it travels from a source at infinity, passes a distance of closest approach, $R$, and returns to infinity. In this section, we compare $\delta$ obtained from full GR in static, spherically symmetric spacetime~\citep[i.e., in Schwarzschild metric][]{carroll97, 2009igr..book.....R, 2009fcgr.book.....S, 2023OJAp....6E..50M} with those derived from the ABN potential with $\beta=2$ \cite{abn96}. In addition, we also compare $\delta$ around a rotating black hole~\citep[i.e., in Kerr metric][]{2005PhRvD..72h3003B, 2006PhRvD..74f3001B} in GR with Mukhopadhyay potential~\cite{bm02}.

%%%%%%%%%%%%%%%%%%%%%%%%%%%%%%%%%%%%%%%%%%%%%%%%%%%%%%%%%%%%%%%%%%%%%%
\subsection{Lensing in a static spherical symmetric spacetime: Schwarzschild vs. ABN}
\label{ssec:schwarschild}
%Gravitational lensing refers to the bending/deflection of light rays as they pass through the gravitational field of a massive object. It is quantified by the total deflection angle, $\delta$, which measures the deviation of the photon's path from its original trajectory as it travels from a source at infinity, passes a distance of closest approach, $R$, and returns to infinity. We compare the deflection angle $\delta$ obtained from full GR \cite{carroll97, 2009igr..book.....R, 2009fcgr.book.....S, 2023OJAp....6E..50M} with those derived from the ABN potential \cite{abn96}.

For the exact GR baseline in the static, spherically symmetric Schwarzschild metric, $\delta$ is obtained by numerically integrating the exact null geodesic equation for a photon. In terms of $\alpha$, the equation for the trajectory in GR is given by \cite{carroll97}:
\begin{equation}
\frac{d^2\alpha}{d\theta^2} + \alpha = 3\alpha^2.
\label{eq:gr_null}
\end{equation}
Clearly this relativistic light bending equation is non-linear. To find the deflection angle, we integrate this equation from the distance of closest approach, where $\alpha_{\max} = 1/R$ (with $d\alpha/d\theta = 0$), out to infinity ($\alpha \to 0$).

For the pseudo-Newtonian approach, we have to solve Eq. (\ref{orbit1}) with the underlying force $F = -dV/dx$ given in \S\ref{sec:peri}. Similar to the GR baseline, the numerical integration is performed using an explicit Runge-Kutta method of order 5(4) (RK45). This is implemented via the \texttt{solve\_ivp} function within the \texttt{SciPy} library \cite{scipy}, using a high-precision variable-step integration with boundary conditions matching the distance of closest approach out to infinity.

\begin{figure}[H]
    \centering
    \includegraphics[width=1\textwidth]{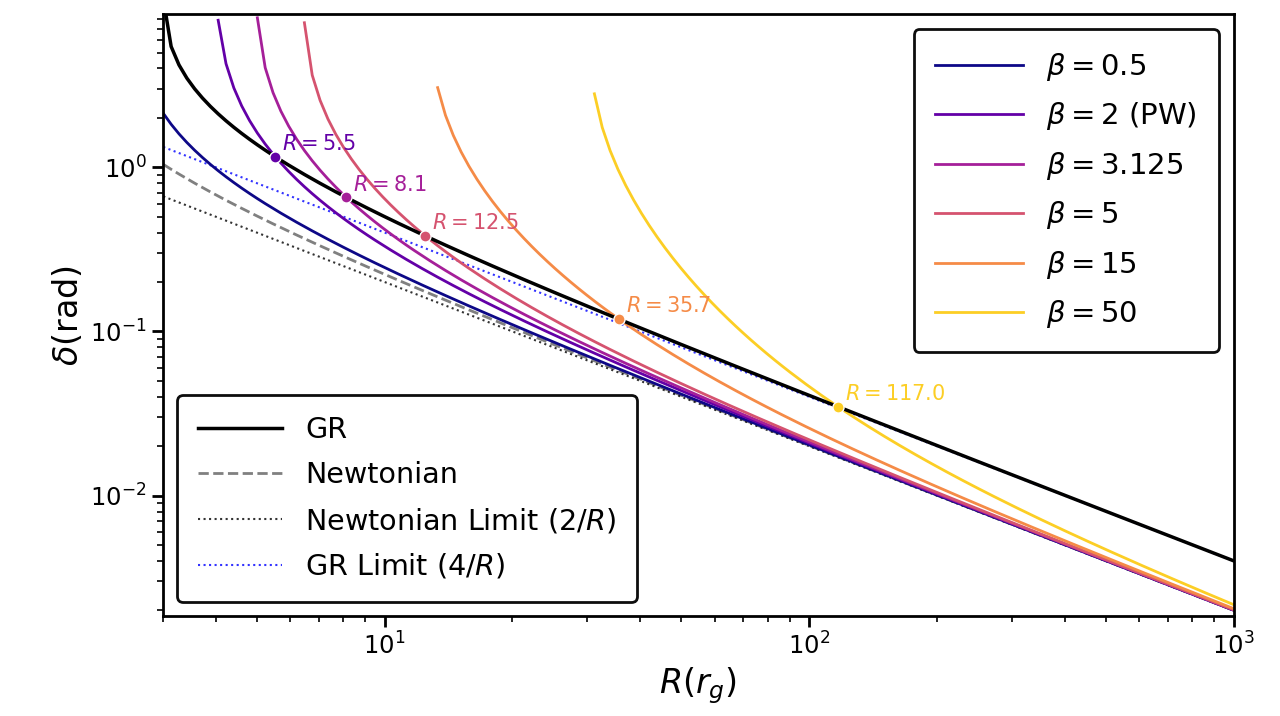}
    \caption{Variation of the deflection angle $\delta$ as a function of radial coordinate $R$ (distance of closest approach) for the ABN potential with different $\beta$. The exact solution from GR (black) is shown for comparison. While there are respective specific $R$ for each $\beta$ where ABN potential matches with GR, there is no generic trend. Also for standard values like $\beta=2,~3$, ABN based results diverge significantly from the GR predictions in the weak and strong field regimes.}
    \label{fig:beta_lensing}
\end{figure}

Fig.~\ref{fig:beta_lensing} shows how $\delta$ varies with $R$ for various values of $\beta$ in the ABN potential \cite{abn96}. Interestingly, the behavior here contrasts sharply with other physical phenomena discussed above. The standard PW potential (i.e., $\beta=2$ in the ABN potential) \cite{pw80}, which suffices very nicely for accretion disk dynamics, significantly under-predicts the deflection angle compared to GR. Similarly $\beta \approx 3$, which leads to the exact perihelion precession of Mercury, also fails to capture the correct lensing behavior. As seen in Fig. \ref{fig:beta_lensing}, to match the GR prediction in the weak-field regime (e.g., for the Sun at $R \approx 10^5 r_g$), one would require an absurdly large value of $\beta$ (likely $\beta \gg 1000$). Moreover, different $\beta$-s are required to match ABN results with GR at different $R$-s. Therefore, while other basic GR tests are confirmed by the pseudo-Newtonian formalism, it is not viable for the bending of light.

%%%%%%%%%%%%%%%%%%%%%%%%%%%%%%%%%%%%%%%%%%%%%%%%%%%%%%%%%%%%%%%%%%%%%%

%%%%%%%%%%%%%%%%%%%%%%%%%%%%%%%%%%%%%%%%%%%%%%%%%%%%%%%%%%%%%%%%%%%%%%
\subsection{Lensing in a rotating black hole spacetime: Kerr vs. Mukhopadhyay}
\label{app:kerr}
While the above sections focus on spherically symmetric spacetimes (appropriate for the Sun), astrophysical black holes often possess significant spin which breaks spherical symmetry and introduces frame-dragging.
To evaluate the efficacy of the Mukhopadhyay potential \cite{bm02} in this regime, we  compare its predicted deflection angles against that based on exact null geodesics in the Kerr metric. The exact Kerr trajectories are computed by solving the geodesic equations of motion governing a test particle in the equatorial plane, as detailed in \cite{bm02}.

%%%%%%%%%%%%%%%%%%%%%%%%%%%%%%%%%%%%%%%%%%%%%%%%%%%%%%%%%%%%%%%%%%%%%%
\begin{figure}[H]
    \centering
    \includegraphics[width=1\textwidth]{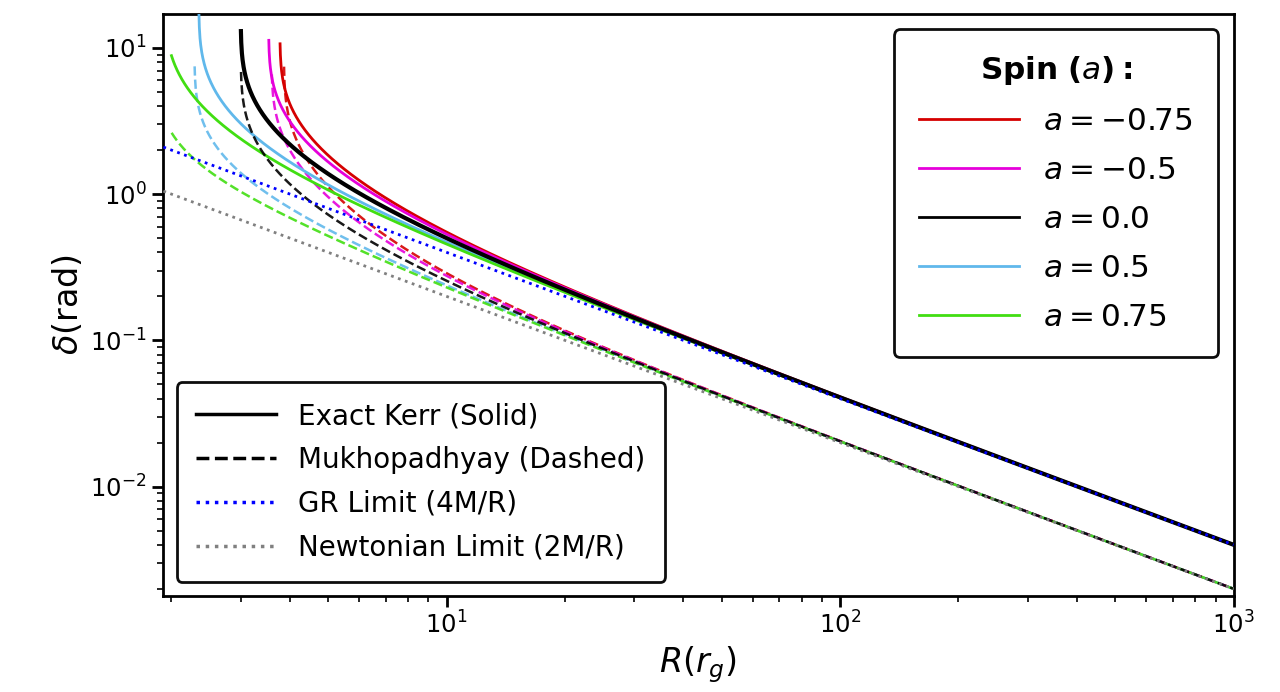}
    \caption{Variation of the deflection angle $\delta$ in equatorial plane as a function of radial coordinate $R$ (distance of closest approach) for the Mukhopadhyay potential with different spin parameters. Solid lines represent the exact Kerr metric predictions; dashed lines correspond to the Mukhopadhyay pseudo-Newtonian potential. The spin parameter $a$ varies from -0.75 to +0.75.}
    \label{fig:kerr_comparison}
\end{figure}

Figure \ref{fig:kerr_comparison} illustrates the results for spin parameters ranging from $a = -0.75$ to $a = +0.75$. We observe the following:

\begin{enumerate}
    \item \textbf{Photon Sphere Asymptote:} As we approach the photon sphere (the vertical asymptote where deflection $\to \infty$), the results with the Mukhopadhyay potential approach the exact Kerr results. Crucially, they both diverge at the \textit{same} radius which is the dimensionless photon radius $x_{ph}$ for a given spin parameter. This confirms that the potential accurately captures the strong-field photon sphere location.
    \item \textbf{Spin Splitting:} The potential correctly reproduces the qualitative splitting of orbits: retrograde ($a<0$) deflection is greater than prograde ($a>0$) deflection for the same impact parameter, consistent with the frame-dragging effect acting against the photon's motion.
    \item \textbf{Weak Field Divergence:} At the large $R$ limit, a clear separation appears. The Kerr results coincide with the weak field GR limit ($4/R$), while the Mukhopadhyay potential results coincide with the Newtonian limit ($2/R$).
\end{enumerate}

%%%%%%%%%%%%%%%%%%%%%%%%%%%%%%%%%%%%%%%%%%%%%%%%%%%%%%%%%%%%%%%%%%%%%%
\section{Conclusion}
\label{sec:conclusion}

General relativity is one of the most important basic building blocks to uncover phenomena in astrophysics and strong field gravity. Three of the classic predictions of GR are Mercury's perihelion precession, bending of light and gravitational redshift which are all repeatedly confirmed by various tests and experiments.

As an alternative to pure GR, researchers often invoke the pseudo-Newtonian approach, capturing important features of GR without going through its mathematical rigor.
This has been well tested in many astrophysical processes successfully. In the present work, we have explored pseudo-Newtonian potentials to possibly uncover classic GR predictions. This helps to assess the broader applicability of the pseudo-Newtonian approach. This also helps to reach the classic GR phenomena and tests to the broader audience who are not expert in GR. 

We find that the while pseudo-Newtonian framework is a very good approach for perihelion precession of planets and gravitational redshift, it does not suffice for gravitational lensing. Moreover, for the former two, the same pseudo-Newtonian model (or model parameter) does not suffice. 
Therefore, while the pseudo-Newtonian potential is a very useful tool to uncover GR phenomena without rigorous geometry, there does not seem to be a unique potential for all purposes. For the accretion disk around a nonrotating black hole, $\beta=2$ is optimal \cite{pw80}. Even that for a rotating black hole, Mukhopadhyay potential \cite{bm02} suffices. For the perihelion precession of Mercury it is $\beta \approx 3$; for gravitational redshift $\beta=0.5$ does; yet for gravitational lensing, no single ``low-$\beta$"  works across all regimes. Adopting the high $\beta$ required for lensing would result in unphysical 
$x_{\rm ms}$ and $x_{\rm mb}$. Nevertheless, far away from the event horizon, all the pseudo-Newtonian models converge to the Newtonian limit, though only the tuned potentials capture the full GR deflection. Similarly, asymptotically the Mukhopadhyay potential based results for deflection get separated from those of well-known GR, as shown in \S\ref{app:kerr}. 

Indeed, it is known that for epicyclic oscillation frequencies none of these $\beta$ seem to be good; one requires a different pseudo-Newtonian approach \cite{nw91, mm03}. 
This is because capturing temporal dynamics, involved with frequencies, requires a different potential than those of spatial dynamics involved with MHD particularly.
Therefore, while in many occasions, pseudo-Newtonian potentials do a nice job to capture correct GR and astrophysical dynamics, there is no unique potential capturing all gravitational physics and astrophysics. More so, some phenomena cannot be captured by the pseudo-Newtonian approach.

Nevertheless, our results can help make the classic GR tests accessible to a wider audience, at least in most of the occasions. 
This will be useful to introduce modified GR effect to understand strong field gravity phenomena by pseudo-Newtonian approach. A better potential, however, may do the job in a broader perspective, what is yet to be searched though.

%%%%%%%%%%%%%%%%%%%%%%%%%%%%%%%%%%%%%%%%%%%%%%%%%%%%%%%%%%%%%%%%%%%%%%
\section*{Acknowledgment}
NG acknowledges financial support from the KVPY fellowship, funded by the Department of Science and Technology (DST), Government of India. 
AKM acknowledges the support from the Start-up Grant IE/CARE-25-0305 provided by the IISc, Bengaluru, India.
This work made use of the \texttt{NumPy} \cite{numpy} and \texttt{Matplotlib} \cite{matplotlib} Python libraries for numerical computations and data visualization.
%%%%%%%%%%%%%%%%%%%%%%%%%%%%%%%%%%%%%%%%%%%%%%%%%%%%%%%%%%%%%%%%%%%%%%
\bibliography{references}

\end{document}